\documentclass{birkjour}   

    \usepackage{hyperref}  
\usepackage{amssymb} 
\usepackage{amsfonts} 
\usepackage{amsmath}

\usepackage{url}
\usepackage[super]{cite}
\usepackage{xcolor}

\usepackage{graphicx} 

 \usepackage{yhmath} 

\usepackage{soul}
\definecolor{TODOcolor}{HTML}{888800}
\definecolor{REDcolor}{HTML}{ff8888}
\definecolor{CYANcolor}{HTML}{00ffff}
\definecolor{olivegreen}{HTML}{BAB86C}
\definecolor{teagreen}{rgb}{0.82, 0.94, 0.75}
\definecolor{lightpurple}{rgb}{0.82, 0.75, 0.94}
\DeclareRobustCommand{\JK}[1]{{\sethlcolor{lightpurple}\hl{#1}}} 

\newcommand{\beq}{\begin{equation}}
\newcommand{\eeq}{\end{equation}}
\newcommand{\beqa}{\begin{eqnarray}}
\newcommand{\eeqa}{\end{eqnarray}}

\newcommand{\we}{\wedge}
\newcommand{\der}{\partial}

\newcommand{\ka}{\varkappa}

\newcommand{\Phib}{\overline{\Phi}}

\newcommand{\what}[1]{\widehat{#1}}

\newcommand{\bx}{{\mathbf{x}}}

\newcommand{\abar}{\bar{a}{}} 
\newcommand{\bbar}{\bar{b}{}} 
 
\newcommand{\omt}{\tilde{\omega}{}}

\DeclareMathOperator{\Tr}{Tr} 

\newcommand{\ugamma}{\underline{\gamma\!}\,}

\newcommand{\abbar}{\bar{\bar{a}}}
\newcommand{\mubbar}{\bar{\bar{\mu}}}

\makeatletter
\renewcommand{\tableofcontents}{}%
\makeatother

\begin{document}

\title[Modifications of Newtonian dynamics from quantum spin connection...]{\huge
Modifications of Newtonian dynamics from higher moments of 
quantum spin connection in precanonical quantum gravity
}


\author[ME Pietrzyk]{\large Monika E. Pietrzyk}
\address{\normalsize    School of Physics and Astronomy, University of Birmingham, \\
  Birmingham B15 2TT, UK 
  }
\email{m.pietrzyk@bham.ac.uk}  
  
\author[VA Kholodnyi, IV Kanatt\v{s}ikov]{\large Valery A. Kholodnyi,$^*$ \mbox{Igor~V.~Kanatt\v{s}ikov$^{\star}$}  }
\address{\normalsize $^*$ Wolfgang-Pauli-Institut, Oskar-Morgenstern-Platz 1, 1090 Vienna, Austria \\ 
Unyxon, Woodforest TX, USA 
\\ $^{\star}$ National Quantum Information Center in Gda\'nsk (KCIK), Gda\'nsk 80-309, Poland 
and IAS-Archimedes Project, 213 Rue de la Soleillette, Saint-Rapha\"{e}l 83700, 
France
} 

\author[J Kozicki]{\large Janek~Kozicki}
\address{\normalsize Faculty of Applied Physics and Mathematics, Gda\'{n}sk University of Technology, 80-233 Gda\'{n}sk, Poland \\
 {Advanced Materials Center, Gda\'{n}sk University of Technology, 80-233 Gda\'{n}sk, Poland } 
}




\begin{abstract} \normalsize 
Building upon previous work that derived an alternative to (galactic) dark matter in the form of  
Modified Newtonian Dynamics (MOND), with a specific theoretical interpolating function, from the motion of a non-relativistic test particle in the gravitational field of a point mass immersed in the non-relativistic static limit of the spin connection foam --- which represents the quantum analogue of Minkowski spacetime within precanonical quantum gravity --- we now show the consequences of using higher moments (third and fourth) of the corresponding geodesic equation with a random spin connection term. These higher moments lead to more general quantum modifications of the Newtonian potential (qMOND potentials expressed in terms of Gauss and Appell hypergeometric functions), more general (steeper) MOND interpolating functions, and a new modification of MOND at low accelerations (mMOND) that features an almost-flat asymptotic rotation curve $\propto r^{-1/18}$, 
which is expected to operate at approximately the same galactic scales as MOND. \\
%
\end{abstract}




\maketitle


\medskip 


\begingroup
\parindent=0pt
\leftskip=0pt
\rightskip=0pt
\parskip=0pt

\centerline{\bfseries\Large Contents} 
\vspace{1em}

\def\tocsectfont{\normalfont} 

\newcommand{\tocentry}[2]{%
    \noindent #1 \dotfill \hspace{3mm} #2 \par
}

\tocentry{\tocsectfont 1\quad \; \;\,\ Introduction}{2}

\tocentry{\tocsectfont 2\quad \; \;\,\ Precanonical quantization, pQG, SCF of quantum Minkowski\\
\hspace*{2em} \;\, spacetime, and its non-relativistic static limit}{3}

\vspace{0.5em}

\tocentry{\tocsectfont 2.1\quad \;\, Quantum states (SCF) of Minkowski spacetime}{6}

\tocentry{\tocsectfont 2.1.1\quad  Non-relativistic static SCF}{7}

\tocentry{\tocsectfont 3\quad \; \;\,\  Quantum modifications of Newtonian potential from \\ 
 \hspace*{2.5em} \,   the static SCF}{8}

\tocentry{\tocsectfont 3.1\quad \;\, qMOND and MOND from the 2nd moment of \\
\hspace*{2.5em} \, the spin connection distribution}{8}

\tocentry{\tocsectfont 3.1.1\quad The rotation curve in the qMOND potential $\boldsymbol{\Phi^{(2)}}$}{10}

\tocentry{\tocsectfont 3.1.2\quad The emergence of MOND}{11}

\tocentry{\tocsectfont 3.2\quad \;\, qMOND from the 4th moment of the spin connection \\
\hspace*{2.5em} \, distribution}{12}

\tocentry{\tocsectfont 3.2.1\quad  The rotation curve in $\boldsymbol{\Phi^{(4)}}$ potential}{13}

\tocentry{\tocsectfont 3.3\quad\;\, MOND from the 4th moment of the spin connection \\
\hspace*{2.5em} \, distribution}{14}

\tocentry{\tocsectfont 3.4\quad\;\, qMOND and a generalized MOND from the 3rd moment of \\
\hspace*{2.5em} \, the spin connection distribution}{17}

\tocentry{\tocsectfont 3.5\quad \;\, The rotation curve in $\boldsymbol{\Phi^{(3)}}$ potential}{17}

\tocentry{\tocsectfont 3.5.1\quad The emergence of a modified MOND (mMOND)}{19}

\vspace{0.5em} 

\tocentry{\tocsectfont 4\qquad \, Conclusions}{22}

\vspace{0.5em}
\tocentry{\tocsectfont Acknowledgments}{25}
\tocentry{\tocsectfont References}{26}

\endgroup
\vspace{1em}

\large 

\section{Introduction}

In recent papers \cite{epl25,dice24,mg17,bial22}, it has been argued that the effects of dark energy (manifested as the cosmological constant $\Lambda$) and dark matter (in the form of Modified Newtonian Dynamics, MOND \cite{mond1,mond2,mond3,mond4}, at low accelerations at or below the Milgrom's acceleration 
$a_0 \sim \sqrt{\Lambda}$) can be understood as manifestations of quantum spin connection foam (SCF). This SCF concept, introduced within precanonical quantum gravity (pQG) \cite{ikv1,ikv2,ikv3,ikv4,ikv5}, describes the quantum geometry of spacetime via quantum fluctuations of the spin connection. 
This description is naturally framed in terms of precanonical wave functions and transition amplitudes defined on the fibered space coordinatized by the spin connection components $(\omega_\alpha^{IJ})$ and spacetime coordinates $(x^\alpha)$. These amplitudes satisfy the 
analogue of the Schr\"odinger equation (see eq. (\ref{psegrav}) below) on this fibered space, which is obtained through the precanonical quantization procedure (see below). 

The results of these papers rest upon two pillars: 
(i) the possibility of finding a solution of that equation corresponding to quantum states (or SCF) of Minkowski spacetime \cite{ik-mink} and (ii) a consideration of test particles moving in the gravitational field of a massive body while simultaneously subject to the spin connection fluctuations of the SCF of quantum Minkowski spacetime.\cite{epl25,dice24,mg17} 
A modification of Newtonian dynamics follows from the quantum average of the square of the corresponding geodesic equation. This equation includes a random connection, the distribution function of which is explicitly given by the quantum gravitational wave function obtained (within pQG) in the non-relativistic static limit of the SCF. 

However, the variance alone of the acceleration in the geodesic equation with a random connection does not provide a complete description of modified dynamics. Consequently, this paper explores the modifications of Newtonian dynamics that follow from considering higher-order moments of the geodesic equation, specifically for a test particle in the field of a fixed point mass immersed in the non-relativistic static limit of the spin connection foam of Minkowski spacetime.

We will proceed as follows. In Section 2, we briefly overview the framework of precanonical quantization and precanonical quantum gravity (pQG). We then describe the simplest solution of pQG corresponding to the quantum state, or SCF of Minkowski spacetime, and its non-relativistic static limit.
In Section 3, we obtain quantum modifications of the Newtonian potential by considering the second, third, and fourth moments of the geodesic equation with a random spin connection accounting for quantum fluctuations of the spin connection in SCF. We then discuss the emergence of MOND and its extension, mMOND, in non-inertial reference frames defined by the mean-field accelerations arising from quantum fluctuations in SCF, which lead to the growing asymptotes of those potentials and give rise to the Milgromian acceleration scale $a_0$. 
Finally, Section 4 presents our conclusions and outlook.




 
\section{Precanonical~quantization,~pQG,~SCF~of~quantum~Minkowski  
spacetime, and its non-relativistic static limit}


Let us recall that pQG is derived from the precanonical quantization of the Palatini formulation of vielbein gravity. This precanonical quantization procedure \cite{ik1,ik2,ik3,ik4,ik5} 
relies on quantizing the analogues of Poisson \cite{mybr1,mybr2,mybr3,ik5}
 or Dirac \cite{mydirac} brackets of differential forms obtained within the De Donder-Weyl (DDW) Hamiltonian formulation of field theory \cite{kastrup}, 
 a method that treats all spacetime variables on an equal footing.
In the case of the vielbein Palatini formulation of general relativity, the DDW Hamiltonian formulation leads to constraints. Their subsequent analysis, following the generalized Dirac procedure \cite{mydirac}, results in a theory formulated in terms of a wave function, $\Psi (\omega, x)$, defined on the space of spin connection components ($\omega_\alpha^{IJ}$) and spacetime variables ($x^\alpha$). 
The corresponding spacetime-symmetric precanonical generalization of the canonical Schr\"odinger equation -- where the time derivative of the wave function(al) is given by the operator of the canonical Hamiltonian function(al) -- takes the form of the Dirac-type operator equal to the operator $\hat{H}$ of the analog of the Hamiltonian within the DDW theory, both acting on the precanonical wave function.

Dimensional considerations introduce the Planck constant $\hbar$  in front of the time derivative in the canonical Schr\"odinger equation. Similarly, because 
 the DDW Hamiltonian function (the analogue of the Hamiltonian within the DDW theory) has the dimension of energy density, we must introduce the factor $\hbar \varkappa $ in front of the Dirac operator, 
 %
 %
 where the parameter $\ka$ has the dimension of the inverse spatial volume. Thus, the most general precanonical Schr\"odinger equation (pSE) assumes the form:
\beq
 i\hbar\ka \what{\not\hspace*{-0.2em}\nabla} \Psi = \what{H} \Psi  \, .  
\eeq
The hat over the Dirac operator signifies that in the context of quantum gravity, based on both the Einsteinian formulation (in vielbein \cite{ikv1,ikv2,ikv3,ikv4,ikv5} and metric \cite{ikm1,ikm2,ikm3,ikm4} variables) and its teleparallel equivalent \cite{tp1,tp2}, either curved-spacetime Dirac matrices or spin connection coefficients are promoted to differential operators.

The relation between precanonical quantization and the standard Quantum Field Theory (QFT) in the functional Schr\"odinger representation has been established for scalar fields \cite{iks1,iks2,iksc1,iksc2,iksc3} and for Yang-Mills fields \cite{iky1,iky3}. The conclusion is that the standard QFT can be obtained from the equations resulting from precanonical quantization in the case of infinitesimal $1/\ka$. This is required to invert the quantization map from differential forms to Clifford algebra elements -- a key part of the precanonical quantization procedure based on the quantization of Poisson-Gerstenhaber brackets of differential forms. 
In particular (in flat (3+1)-dimensional spacetime), the inverse quantization map  takes the form 
\beq
 \frac{1}{\ka} \gamma^0 \mapsto   dx^1\we dx^2 \we dx^3  \, .
\eeq 
This map allows for a transition from the description in terms of Clifford algebra-valued precanonical wave functions to the description in terms of Schr\"odinger wave functionals.

Now, the application of the precanonical quantization procedure to the Einstein-Hilbert-Palatini 
action in vielbein variables that also includes the cosmological term, leads to the following results:
\cite{ikv1,ikv2,ikv3,ikv4,ikv5}  
\begin{itemize}
\item The operator representation of 
vielbein components:
\beq \label{ebiop} 
\hat{e}{}^\alpha_I   = -  8\pi i \hbar G \varkappa   \ugamma{}^{J}\frac{\der}{\der \omega_{\alpha}^{IJ}} ,
\eeq
where $\ugamma^I$-s satisfy the flat spacetime Clifford algebra:
 $\ugamma^I \ugamma^J + \ugamma^J \ugamma^I = 2 \eta^{IJ}$. 
 \item  The operator representation of contravariant components of the metric tensor 
\beq \label{gop}
\hat{g}{}^{\alpha\beta} = - 64 \pi^2 \hbar^2 G^2 \varkappa^2 \eta^{IK} \eta^{JL} \frac{\der^2}{\der {\omega_\alpha^{IJ}} \der {\omega_\beta^{KL}}} .    
\eeq

\item The explicit form of the precanonical Schr\"odinger  equation (pSE) for pure quantum gravity without matter 
\beq  \label{psegrav}
\ugamma{}^{IJ}  \frac{\der}{\der \omega_{\alpha}^{IJ}}   
   \Big ( \der_\alpha + \frac{1}{4} \omega_{\alpha}^{KL}\ugamma{}_{KL}\!\! \stackrel{{\leftrightarrow}}{\vee}  
 -\,
\omega_{\alpha M}^{K}\omega_{\beta}^{ML}  \frac{\der}{\der \omega_{\beta}^{KL}} \Big)  
 \Psi      
 -  \lambda \Psi  = 0 \, ,  
\eeq
where $\ugamma{}^{IJ}=\frac12 (\ugamma^I\ugamma^J -  \ugamma^J\ugamma^I)$,   
  $\Psi = \Psi (\omega,x)$ is the Clifford-algebra-valued precanonical wave function 
  on the bundle of spin connection components over a differentiable spacetime manifold, 
\beq \label{cprod}
{\ugamma}{}^{IJ} \stackrel{\leftrightarrow}{\vee} \Psi 
  = \frac12 \left (\ugamma{}^{IJ} \Psi - \Psi \ugamma{}^{IJ}  \right) , 
\eeq
and 
\beq \label{lam}
\lambda = \frac{\Lambda}{ (8\pi\hbar G \varkappa)^2}    
\eeq 
is a dimensionless combination of the cosmological constant $\Lambda$, the parameter $\varkappa$, and the fundamental constants $\hbar$ and  $G$. The numerical value of ${\lambda}$ is determined by the ordering of operators in Eq. (\ref{psegrav}) and depends only on the dimensionality of spacetime, which specifies both the dimensionality of the Clifford algebra and the number of components of the spin connection.
\item The scalar product of Clifford-algebra-valued wave functions is defined as:
\beq \label{scprod}
\left\langle \Phi | \Psi \right\rangle
:=\Tr\! \int \Phib  \what{[d\omega]}{} \Psi \, .  
\eeq
Here, ${\bar{\Psi}}$ defined as ${\bar{\Psi} := \gamma^0 \Psi^\dagger \gamma^0}$ (using the standard Dirac matrix $\ugamma^0= \beta$) is a generalization of the Dirac conjugate for spinors to a Clifford-valued $\Psi$  
and $\what{[d\omega]}$ is the invariant measure on the space of spin connection components:
\beq \label{measop}
\what{[d\omega]}
\sim \hat{\mathfrak e}{}^{-6}\prod_{\mu, I,J} d \omega_\mu^{IJ} ,
\eeq
with the operator $\hat{\mathfrak e}{}^{-1}$ constructed from the operators introduced in Eq.~({\ref{ebiop}}).
\end{itemize}

\subsection{Quantum states (SCF) of Minkowski spacetime}

For the states of the quantum analogue of Minkowski spacetime in Cartesian coordinates, where 
$\omega_{\alpha}^{IJ} = 0$ and $\Lambda=0$,  
the requirement $\langle \hat{g}^{\alpha\beta} \rangle = \eta^{\alpha\beta}$ leads to the equation 
\beq\label{mi2}
\hat{g}^{\alpha\beta} \Psi = - (8\pi \hbar G \varkappa)^2 \eta^{IK} \eta^{JL} \frac{\partial}{\partial \omega_\alpha^{IJ}} \frac{\partial}{\partial \omega_\beta^{KL}} \Psi =  \eta^{\alpha\beta} \Psi \, .
\eeq
In this case, the pSE in (\ref{psegrav}) significantly simplifies and takes the form 
\beq \label{mi1}
\ugamma^I\hat{e}^\alpha_I \der_\alpha \Psi = 0, 
\eeq
where the operator  $\hat{e}^\alpha_I$ is given by (\ref{ebiop}). 
The square of this simplified pSE in (\ref{mi1}) is: 
\beq \label{gdd}
\hat{g}^{\alpha\beta}  \der_\alpha \der_\beta \Psi = 0. 
\eeq 
This equation, together with eq. (\ref{mi2}), then leads to the conclusion that the quantum states of Minkowski spacetime consist of 
(i) light-like modes propagating along the spacetime base according to the d'Alembert equation, 
\beq \label{mi3}
\eta^{\alpha\beta} \der_\alpha \der_\beta \Psi = 0 , 
\eeq 
and (ii) massive modes in four $(3+3)-$dimensional subspaces of $\omega_\alpha^{IJ}$ (in $(3+1)$ spacetime dimensions), which satisfy ultra-hyperbolic Klein-Gordon equations (\ref{mi2}). 
 The mass of these modes defines the characteristic range 
\beq \label{astar}
a_* = 8\pi \hbar G \varkappa 
\eeq
in the space of spin connection components. Physically, it defines the scale of accelerations below which the quantum fluctuations of the spin connection invalidate the classical notion of an inertial reference frame.
A comparison with (\ref{psegrav}) shows that $a_*$ is related to the cosmological constant $\Lambda $ and the dimensionless constant $\lambda $ by 
   \beq \label{astarla}
a_* = \sqrt{\frac{\Lambda}{\lambda}}\, . 
\eeq  
 This relationship is reminiscent of the relationship between the Milgromian acceleration scale, $a_0$, 
 and the square root of the cosmological constant found in Modified Newtonian Dynamics (MOND  \cite{mond1,mond2,mond3,mond4}) in the context of flat galaxy rotation curves.  While this relationship represents a significant theoretical challenge for MOND, in pQG it appears as an intrinsic feature. 

Moreover, previous work \cite{epl25,mg17,dice24,bial22} has established that pQG offers a theoretical basis for explaining the small observed values of $\Lambda$ and $a_*$. Specifically, the characteristic acceleration scale is estimated in geometrized units as
\beq \label{astarnum}
a_* \sim 10^{-27}\ \mathrm{m}{}^{-1} \, .
\eeq 
However, the current theoretical error of this estimation spans several orders of magnitude due to uncertainties in estimating the scale of the parameter $\varkappa$. Crucially, the estimated value of $\varkappa$ corresponds to the hadronic scale of the mass gap in the quantum pure Yang-Mills sector of the Standard Model \cite{my-ymmg}. This naturally situates the characteristic scales of pQG, such as $a_*$ and $\Lambda$, in the cosmological domain rather than the Planckian domain, as is conventionally assumed in most theories of quantum gravity.

\subsubsection{Non-relativistic static SCF}

In the non-relativistic approximation and for a static spacetime, where the only non-vanishing components of the spin connection are $\omega_0^{i0}$ (denoted $\tilde{\omega}{}^i$),   which coincide with $\Gamma^i_{00}$, 
the equations (\ref{mi2}), (\ref{mi3}) defining the state of the non-relativistic static
SCF read 
\begin{align} \label{mo20a}
\eta^{ij} \der_{\tilde{\omega}{}^i} \der_{\tilde{\omega}{}^j} \Psi  
+  \frac{1}{a_*^2} 
\Psi &= 0 \, , \\    
\eta^{ij}\der_i\der_j \Psi &= 0 \, ,  \label{mo20p}
\end{align}
where $\eta^{ij}= - \delta^{ij}$ 
in our signature of choice, and the indices $i, j$ run 
over $1,2,3$. These two coupled equations with explicitly separated spin connection and space variables  define the quantum state $\Psi (\tilde{\omega}^i, x^i)$: the first is the modified Helmholz equation in 3 dimensional space of the spin connection components ($\tilde{\omega}^i$), and the second is the Laplace equation ensuring the static condition within the spatial base manifold ($x^i$).

The  ground state, $\Psi_0$, which is in the form of the Yukawa potential, is given by  
\beq \label{mo22}
\Psi_0 (\tilde{\omega}{}) 
 = 
   \frac{1}{\sqrt{2\pi a_*}}  \frac{ e^{-\tilde{\omega}/a_* } }{\tilde{\omega}} 
\, , 
\eeq 
where $\tilde{\omega} = \sqrt{{\delta_{ij}\tilde{\omega}{}^i \tilde{\omega}{}^j}}$. 
This function solves (\ref{mo20a}) and is normalized as follows 
\beq
\langle \Psi_0 |\Psi_0 \rangle = 
\int\! d {\tilde{\omega}{}^1}\! \int\!  d {\tilde{\omega}{}^2} \! \int\! d{\tilde{\omega}{}^3} \, 
 |\Psi_0(\tilde{\omega}{}^i)|^2 = 
 4 \pi  \int_0^\infty\! d \tilde{\omega} \, \tilde{\omega}{}^2
 |\Psi_0(\tilde{\omega}{}^i)|^2 
  = 1\,  .  
\eeq
Due to the spherical symmetry of the ground state, the mean value of the spin connection 
is zero, 
 \beq  \label{omi}
 \langle \tilde{\omega}{}^i \rangle = 
\int\! d {\tilde{\omega}{}^1}\! \int\!  d {\tilde{\omega}{}^2} \! \int\! d{\tilde{\omega}{}^3} \, 
 \tilde{\omega}{}^i  |\Psi_0(\tilde{\omega}{}^i)|^2  = 0  \, . 
 \eeq
Similarly, all odd moments are vanishing, meaning 
$\langle \tilde{\omega}{}^{2k}\tilde{\omega}{}^i \rangle = 0$, where $k$ is a positive integer. 
However, the variance,  
\beq
\label{avom2}
\langle \tilde{\omega}{}^i \tilde{\omega}{}_i \rangle 
= 4 \pi  \int_0^\infty\! d \tilde{\omega} \, \tilde{\omega}{}^4 \, |\Psi_0|^2 
= \frac{1}{2} a_*^2 \, ,  
\eeq
and the tensor second moments,  
\beq \label{omij}
\langle \tilde{\omega}{}^i \tilde{\omega}{}^j \rangle 
= \int\! d {\tilde{\omega}{}^1}\! \int\!  d {\tilde{\omega}{}^2} \! \int\! d{\tilde{\omega}{}^3} \, 
 \tilde{\omega}{}^i \tilde{\omega}{}^j |\Psi_0(\tilde{\omega}{}^i)|^2
= \frac16 \delta^{ij} a_*^2 \, ,
\eeq
are not vanishing, nor are higher even central moments, such as the fourth moment 
\beq 
\label{avom4}
\langle (\tilde{\omega}{}^i \tilde{\omega}{}_i)^2 \rangle 
= 4 \pi  \int_0^\infty\! d \tilde{\omega} \, \tilde{\omega}{}^6 \, |\Psi_0|^2 
=  \frac{3}{2} a_*^4  
\, , 
\eeq
and arbitrary even moments defined as follows: 
\beq
\label{avom2k}
\langle (\tilde{\omega}{}^i \tilde{\omega}{}_i)^{k} \rangle 
= 4 \pi  \int_0^\infty\! d \tilde{\omega} \, \tilde{\omega}{}^{2k+2} \, |\Psi_0|^2 
=  \frac{(2k)!}{2^{2k}} a_*^{2k} .   
\eeq

\section{Quantum modifications of Newtonian potential from the  static  SCF}

In this section, we study the influence of spin connection fluctuations in the  non-relativistic static approximation of the Spin Connection Foam (SCF) of Minkowski spacetime on the motion of a non-relativistic test particle in the gravitational field of a point mass $M$ fixed at the origin. This motion is described by the geodesic equation with a random spin connection $\tilde{\omega}{}^i$ whose probabilistic properties are defined by the quantum gravitational precanonical wave function of non-relativistic static SCF described in the previous section, namely:\cite{epl25,dice24,mg17}
\beq  \label{mo3a}
\ddot{x}{}^i   
=  - \mu \frac{x^i}{r^3} - \tilde{\omega}{}^i \, , 
\eeq
where $\mu = GM$ denotes the standard gravitational parameter.

\subsection{qMOND and MOND from the 2nd moment of the spin connection distribution}  \label{sec31}

By averaging the square of (\ref{mo3a}) the following quantum gravitational 
modification of Newton's law, or qMOND, is obtained \cite{epl25,dice24,mg17} 
\beq \label{agener}
|\ddot{\bx}| = \left( \frac{\mu^2}{r^4} + \bar{a}^2 \right)^{1/2}, 
\eeq
where the notation $\langle \tilde{\omega}^2 \rangle = \bar{a}^2 = a_*^2/2$ is used. 

The qMOND law can be expressed in vector form 
\beq  \label{vect}
 \ddot{\mathbf{x}} = -\nabla \Phi,
\eeq 
where the qMOND potential $\Phi^{(2)}(r)$ derived from (\ref{agener}) is expressed in terms of the Gauss hypergeometric function 
\beq \label{phi2}
\Phi^{(2)}(r) = -\frac{\mu}{r} \, {}_2F_1\left(-\frac{1}{2}, -\frac{1}{4}; \frac{3}{4}; -\frac{\bar{a}^2 r^4}{\mu^2}\right)\,  . 
\end{equation} 
This potential behaves like the Newtonian potential, $-\mu/r$, at small $r$, but at very large $r$, it features an asymptotic linear behavior, $\bar{a}r$. The slope of this asymptote -- which manifests quantum randomness of the spin connection in the surrounding SCF -- is extremely small, as its scale is related to the cosmological constant (see Eq. (\ref{astarla})). Figure 1 illustrates this behavior using a slope ($\abar = 0.01$) exaggerated by many orders of magnitude compared to the physical one given by Eq. (\ref{astarnum}) 
and relative to the scales of $\mu$ for known massive formations in the universe. 



\begin{figure}[t]
\centering
        \includegraphics[width=0.8\linewidth]{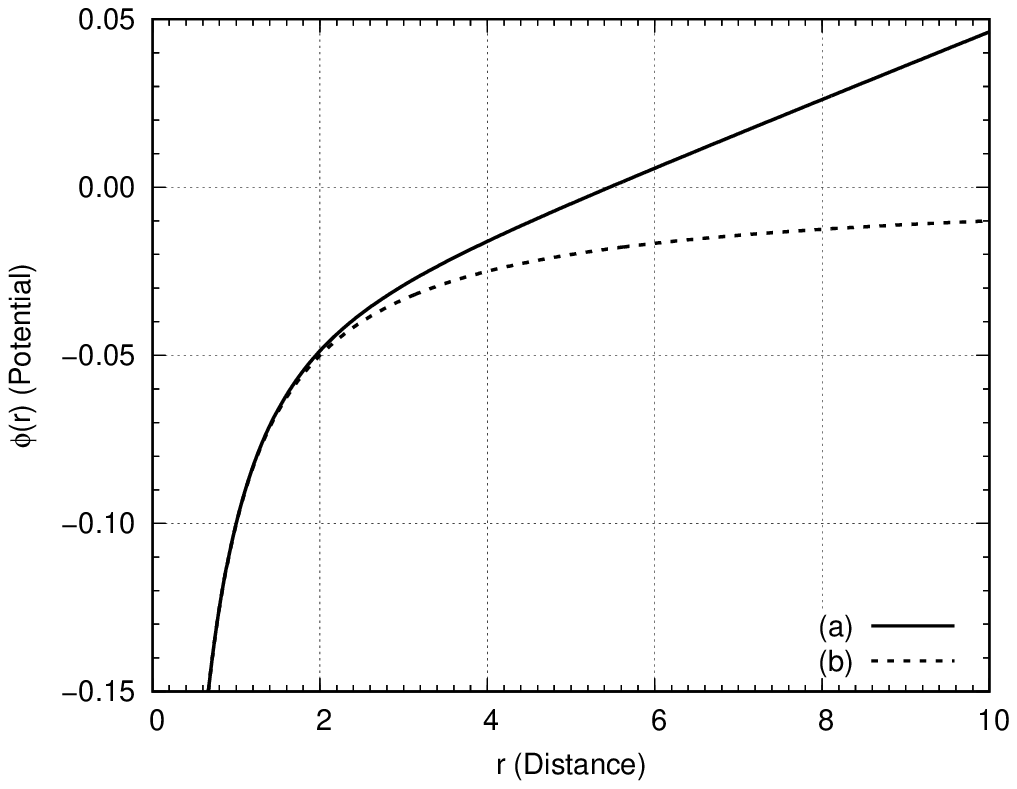}
        \label{fig1} 
        \caption{
         The qMOND potential $\Phi^{(2)}(r)$ for $\mu=0.1$, ${\abar=0.01}$: 
        (a) 
        $\Phi^{(2)}(r)$;  
        (b) 
           $\Phi_{\mathrm{Newton}}(r)=-\frac{\mu}{r}$.}
\end{figure}

Note that the potential in Eq. (\ref{phi2}) is defined up to a constant. Moreover, the hypergeometric series 
defining the Gauss hypergeometric function in (\ref{phi2}) converges only at $r< \sqrt{\mu/\abar}$ so that 
the potential beyond this convergence radius should be defined by analytic continuation. Alternatively, one 
can find an expression of the potential, up to a constant, which converges exactly outside the radius of convergence of the previous expression, namely:   
\beq \label{phi2a}
\Phi^{(2)}(r){}' = \abar {r} \, {}_2F_1\left(-\frac{1}{2}, -\frac{1}{4}; \frac{3}{4}; -\frac{\mu^2}{\bar{a}^2 r^4}\right)\,  . 
\eeq 
Moreover, the constant offset between the two forms of ${\Phi^{(2)} (r)}$, $ \Delta \Phi = \Phi^{(2)}(r)' - \Phi^{(2)}(r)$, can be calculated explicitly \cite{inprep-l}:
\beq \label{dephi}
\Delta \Phi = 2 \left( \frac{\Gamma\left(\frac34 \right)^2}{\sqrt{\pi}} \right) \sqrt{\mu\abar}
\approx 1.694 \sqrt{\mu\abar} \, .
\eeq
In fact, using spectral properties of the operator $r\frac{d}{dr}$\cite{khol}, a single expression for the potential ${\Phi^{(2)} (r)}$ in terms of an Appell function defined for all $r>0$ can be constructed \cite{inprep-l}.


\subsubsection{The rotation curve in the qMOND potential $\boldsymbol{\Phi^{(2)}}$} 

 The orbital velocity of a test particle 
 in the potential $\Phi^{(2)}$ is now given by   
\beq \label{vr}
v(r) = \left( \frac{\mu^2}{r^2} + \abar^2 r^2  \right)^{1/4}  . 
\eeq
This function has a minimum at
\beq \label{rm}
r_m = \sqrt{ \frac{\mu}{\abar}} \, , 
\eeq
where the rotation velocity is 
\beq \label{vm}
v_m := v(r_m) =  \sqrt[4]{2\abar \mu}  \, . 
\eeq  
In the vicinity of this minimum, the rotation curve, $v(r)$, can be approximated by 
the Taylor expansion   
\begin{align}\label{vr1}
v(r) & \approx 
 v_m  +  \frac{\abar^2}{v_m^3} (r - r_m)^2 + O ((r-r_m)^3)   \\
\label{vr2}
&\approx \sqrt[4]{2\abar \mu} + 
 \sqrt[4]{\frac{\abar^5}{8 \mu^3}}
(r - r_m)^2 + O ((r-r_m)^3)  \, .  
\end{align}
This expansion represents a very flat parabola around $r=r_m$. 
This flatness occurs because the scale of the acceleration parameter
 $\abar$ is related to the 
cosmological scale, whereas the scale of $\mu$ (half the Schwarzschild radius of the mass $M$) is many orders of magnitude  smaller.    The smallness of the coefficient ${\sqrt[4]{\abar^5/\mu^3}}$ ensures the rotation 
curve around $r_m$ is approximately flat. 
Figure 2 shows the rotation curve $v(r)$ around the minimum $r_m$ compared with the Keplerian orbital velocity. 


\begin{figure}[t]
\centering
        \includegraphics[width=0.8\linewidth]{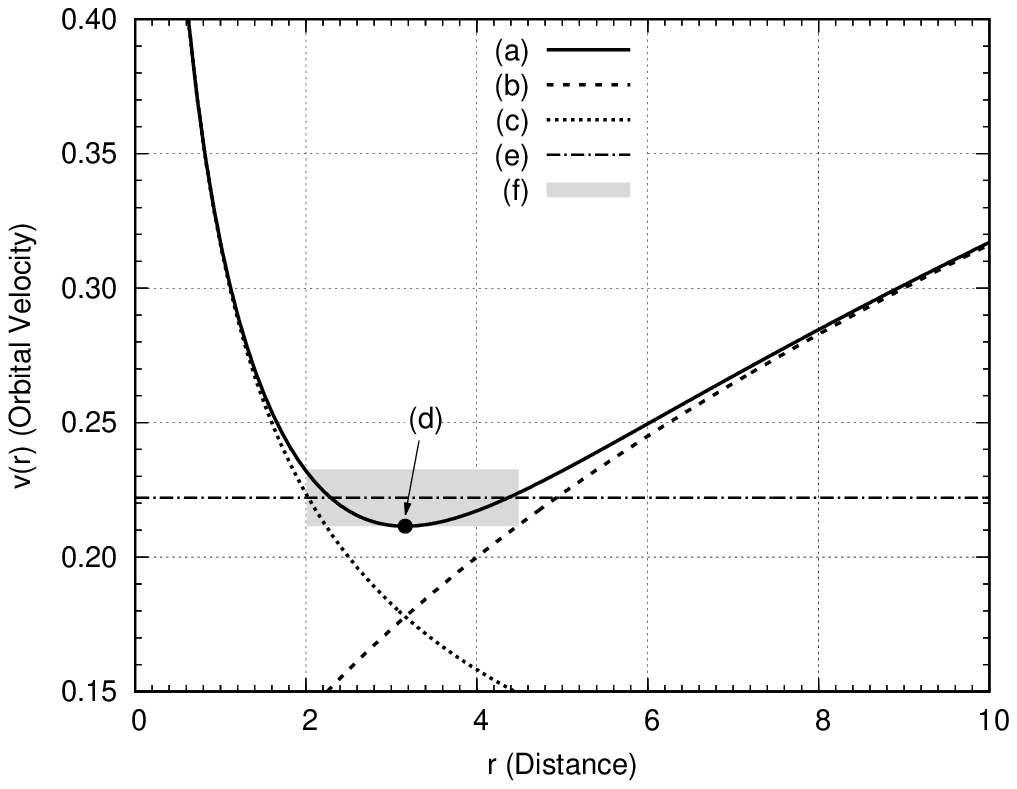}
        \label{fig2} 
        \caption{{Rotation curve $v(r)$ in the qMOND potential $\Phi^{(2)}(r)$ for ${\mu=0.1}$, ${\abar=0.01}$:
        (a) The calculated orbital velocity curve, $v(r)$;
        (b) The asymptotic linear velocity, $v_{\mathrm{asymp}}(r)=\sqrt{\bar{a}r}$;
        (c) The Newtonian (Keplerian) curve, $v_{\mathrm{Kepler}}(r)=\sqrt{{\mu}/{r}}$;
        (d) The velocity minimum, $v_{m}=\sqrt[4]{2 \bar{a} \mu}$, located at radius $r_{m}=\sqrt{{\mu}/{\bar{a}}}$;
	    (e) The line representing the minimum velocity plus 5\%, $v_m + 5\%$;
        (f) The flat rotation curve band, with a 10\% error margin.}
        }
\end{figure}


\subsubsection{The emergence of MOND} \label{sec311}

The mechanism of emergence of MOND from qMOND was found in 
 earlier papers \cite{epl25,dice24,mg17}.  
The core idea is that quantum fluctuations of the spin connection lead to a mean field of spin connection, which is identified with the standard deviation $\abar = \sqrt{\langle \omt^2 \rangle}$. This mean field is manifested in the linear asymptote of the qMOND gravitational potential $\Phi^{(2)}$. This finding makes it reasonable to perform a transformation of the dynamics into the non-inertial reference frame of the global mean field of accelerations, $\abar$, which itself emerges from fluctuations of the non-relativistic static SCF.  This transformation effectively results in the addition of a fictitious force to the Newtonian force:
\beq \label{aa2}
m|\ddot{\bx}| = \frac{GMm}{r^2} + m\abar \,.
\eeq
Then, by defining the effective acceleration $g= |\ddot{\bx}| -\abar$ and by interchanging the roles of force and acceleration in the general equation (\ref{agener}), we obtain a modified dynamics in the form 
\beq  \label{mmondab}
 \frac{GM}{r^2} =  \sqrt{g^2 + \abar{}^2} - \abar \, , 
\eeq
which is identical to the Modified Newtonian Dynamics (MOND) postulated by Milgrom in 1983 as an alternative to the dark matter hypothesis for describing the flat rotation curves of galaxies \cite{mond1,mond2,mond3,mond4}:
\beq \label{mond0}
\frac{GM}{r^2} = \mu\left( \frac{g}{a_0}\right)  g \, , 
\eeq
provided the Milgromian acceleration scale, $a_0$, is related to the standard deviation $\abar$ as follows: 
\beq  \label{g0}
a_0 = 2 \abar \, . 
\eeq
This derivation yields the specific, theoretically derived interpolating function (IF) $\mu(x)$ 
\beq  \label{mux}
\mu(x) = \mu{}^{(2)}(x)
 = \frac{1}{2x} \left(\sqrt{4x^2+1} - 1    \right) ,  
\eeq
which interpolates between the Newtonian dynamics at $ |\ddot{\bx}| \gg a_0$ (where $\mu(x) \to 1$)
and the ``deep-MOND" regime at $ |\ddot{\bx}|  \le a_0$ (where $\mu(x) \to x$).
 
\newcommand{\mvedtotheconclusion}{
Note that the theoretical IF (\ref{mux}) relies on specific physical assumptions, including:
\begin{itemize} 
\item The lack of back-reaction of the central mass $M$ on the state of the SCF. 
\item The central mass remaining rigidly fixed at the origin, despite the quantum fluctuations of spin connection influencing both the test particle and the central body equally -- as required by the Equivalence Principle.
\item The neglect of long-distance correlations in the non-relativistic SCF. These correlations follow from the two-point solutions of (\ref{mo20p}) and may lead to an additional correlational contribution to the force between the test particle and the central mass -- again, due to the Equivalence Principle.
\end{itemize}
The work on lifting those assumptions is currently in progress. Instead, this paper concentrates on obtaining new interpolating functions from the higher-order moments of equation ($\ref{mo3a}$).
}

\subsection{qMOND from the 4th moment of the spin connection distribution}

 Obviously, the second moment of eq. (\ref{mo3a}) does not contain all the information about the dynamics of the system immersed in the SCF. It is, therefore, of interest to study the consequences of averaging the higher degrees of fluctuation.

Let us consider the average of the 4th power of (\ref{mo3a}). Taking into account that 
$\langle \tilde{\omega}{}^i\rangle = 0$, we obtain 
\beq \label{ddotx4}
|\ddot{\bx}|^4 = \left \langle  
\frac{\mu^4}{r^8} + 2\frac{\mu^2}{r^4}\abar^2 
+ 4\frac{\mu^2}{r^6} x^ix^j\tilde{\omega}{}_i \tilde{\omega}{}_j + \tilde{\omega}^4 
\right \rangle  \, .
\eeq
For the mixed moment, $\langle  x^ix^j\tilde{\omega}{}_i \tilde{\omega}{}_j \rangle$, 
from (\ref{omij}) we obtain 
\begin{align} \label{mixed4}
\begin{split}
\left \langle
x^ix^j\tilde{\omega}{}_i \tilde{\omega}{}_j  \right \rangle 
= \frac13 r^2 \abar^2  . 
\end{split}
\end{align}
Therefore, the magnitude of the averaged acceleration is
\beq \label{ddot4}
|\ddot{\bx}| = \left ( \frac{\mu^4}{r^8} + \frac{10}{3}  \frac{\abar^2 \mu^2}{r^4} + \bbar{}^4 \right)^{1/4} ,  
\eeq
where 
$\bbar{}^4 = \langle \tilde{\omega}{}^4\rangle = 6\abar^4$ (see ($\ref{avom4}$)) 
denotes the fourth moment of spin connection. 

By integrating (\ref{ddot4}), one obtains the quantum-corrected Newtonian potential, which then allows us to write the modified dynamics law (\ref{ddot4}) in the vector form (\ref{vect}). This potential is expressed in terms of the two-variable generalization of the Gauss hypergeometric function, known as the Appell function
 $F_1(a;b_1,b_2;c;z_1,z_2)$:\cite{appel13} 
\beq \label{appelldef}
F_1(a; b_1, b_2; c; z_1, z_2) = \sum_{m=0}^{\infty} \sum_{n=0}^{\infty} \frac{(a)_{m+n}(b_1)_m(b_2)_n}{(c)_{m+n} m! n!} z_1^m z_2^n  \, , 
\eeq
where $(q)_k = \frac{\Gamma(q+k)}{\Gamma(q)}$ is the Pochhammer symbol. 

Up to additive constants, the potential derived from (\ref{ddot4}) 
    can be written in two forms \cite{inprep-l}
\begin{align}  
\!\!\Phi^{(4)} (r) 
 =  & - \frac{\mu}{r} \;F_1\!\left(-\frac{1}{4}; -\frac{1}{4}, -\frac{1}{4}; \frac{3}{4};
 - \frac{\alpha\abar^2 r^4}{\mu^2}, - \frac{\alpha^* \abar^2 r^4}{\mu^2} \right) , 
 \label{phi4appellm}
 \\
\!\!\Phi^{(4)}(r){}' 
= & 
\, 6^{\frac{1}{4}}\abar r \;F_1\!\left(-\frac{1}{4}; -\frac{1}{4}, -\frac{1}{4}; \frac{3}{4};-\frac{\mu^2}{\alpha \abar^2 r^4}, -\frac{\mu^2}{\alpha^* \abar^2 r^4} \right),\! 
 \label{phi4appella}   
\end{align}  
where 
$\alpha = \frac13 (5 +i {\sqrt{29}})$ and $\alpha^*$ is its complex conjugate ($\alpha\alpha^*=6$).  
 Similar to the potential ${\Phi^{(2)}}$, these two forms correspond to the converging bivariate hypergeometric series at $r<\sqrt{\overline{b}/\mu}$ and $r>\sqrt{\overline{b}/\mu}$, respectively. 
 In fact, one can find a single expression for the potential ${\Phi^{(4)} (r)}$ in terms of a trivariate Lauricella function defined for all positive real $r$. 
 The details of the derivation of this expression and of Eqs. ($\text{\ref{phi4appellm}}$) and ($\text{\ref{phi4appella}}$) are postponed to a forthcoming technical paper \cite{inprep-l}.
%

 In spite of  the appearance of the complex numbers $\alpha, \alpha^*$,  these potentials 
 in Eqs.  (\ref{phi4appellm}), (\ref{phi4appella}) 
 are real-valued because $F_1$
	is symmetric in its two variables  and the double hypergeometric series 
 (\ref{appelldef}) defining 
	the Appell function $F_1(a; b_1, b_2; c; z_1, z_2) $ for real parameters $(a, b_1, b_2, c)$ has the 
	obvious property 
	$F_1(a; b_1, b_2; c, z_1, z_2)^* = F_1(a; b_1, b_2; c, z_1^*, z_2^*)$. 
	
Analogous to ${\Phi^{(2)}}$, the potential ${\Phi^{(4)}}$ exhibits Newtonian behavior, $-\mu/r$, at small $r$ and a linear asymptote, $\bar{b}r$, at large distances $r$, the far-field slope of $\Phi^{(4)}$ is higher than that of $\Phi^{(2)}$, specifically: $\bar{b} = \sqrt[4]{6} \bar{a} \approx 1.57 \bar{a}$.

\subsubsection{The rotation curve in $\boldsymbol{\Phi^{(4)}}$ potential}

For the orbital velocity in the field of the qMOND potential $\Phi^{(4)}$, we obtain 
\beq  \label{vorb4} 
|\mathbf{v}|^{(4)} = \left(\frac{\mu^4}{r^4} + \frac{10}{3}\abar^2\mu^2 + \bbar^4r^4 \right)^{1/8} . 
\eeq  
The minimum of this function is located at 
\beq \label{rm4} 
r_m^{(4)}=\sqrt{\frac{\mu}{\bbar}} 
\, .
\eeq\
The orbital velocity at this minimum is 
\beq\label{vmin4} 
 v_m^{(4)} =  \left( 2\mu^2 \bar{b}^2 + \frac{10}{3}\mu^2\bar{a}^2 \right)^{1/8}\,. 
\eeq
Taking into account that $\bbar^2 = \sqrt{6} \abar^2$ (see equation (\ref{avom4})), we obtain 
\begin{align}\label{rm4a} 
r_m^{(4)}  &=\sqrt{\frac{\mu}{\sqrt[4]{6}\abar}} \approx 
  0.8 \sqrt{\frac{\mu}{\abar}} \, , 
\\ 
v_m^{(4)} &=  \left[ \frac{2}{3} \left( 5 + 3\sqrt{6} \right) \right]^{1/8} (\mu \bar{a})^{1/4}
 \approx 1.3 \sqrt[4]{\mu \abar}    \, .\label{vm4a} 
\end{align}  
In the vicinity of the minimum, 
\begin{align} 
\begin{split} \label{v4flat} 
|\mathbf{v}|^{(4)}    
&\approx v_m^{(4)}  + 2\, \frac{\mu\bbar^3}{v_m^7}  (r - r_m)^2 + O ((r-r_m)^3)
   \\
&\approx v_m^{(4)} 
  + 1.21 \sqrt[4]{\frac{\abar^5}{\mu^3}} (r - r_m)^2 + O ((r-r_m)^3)   \,.
  \end{split}
\end{align}

The comparison with the results obtained in section (\ref{sec31}) from the averaged square of equation (\ref{mo3a}) shows that, in the case of quantum-modified dynamics derived from the averaged fourth power of equation (\ref{mo3a}), we observe the following effects of the dynamics based on the 
 fourth moment instead of variance: 
\begin{itemize}
\item $r_m^{(4)}$ is $\approx 20\%$ closer to the gravitating center, 
\item $v_m^{(4)}$ is $\approx 9\%$  higher, 
\item the parabola around $v_m^{(4)}$ is approximately two times steeper. 
\end{itemize}

Figure 3 presents the rotation curve $v^{(4)}(r)$ (derived from the qMOND potential $\Phi^{(4)}$) compared against the previous $v(r)$ (derived from $\Phi^{(2)}$) and the standard Keplerian rotation curve.


\begin{figure}[t]
\centering
        \includegraphics[width=0.8\linewidth]{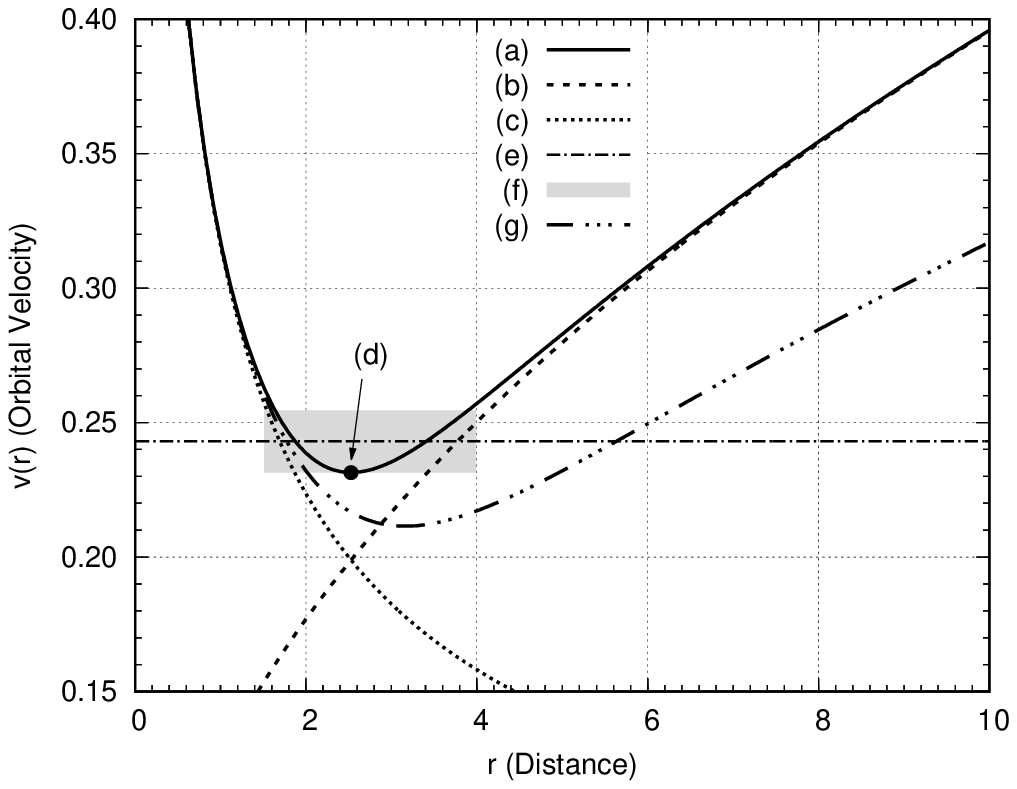}
        \label{fig:vorb4}
        \caption{{Rotation curve $v^{(4)}(r)$ derived from the $\Phi^{(4)}$ potential for $\mu=0.1$ and $\bar{a}=0.01$:
        (a) The calculated orbital velocity curve, $v^{(4)}(r)$;
        (b) The asymptotic linear velocity, $v^{(4)}_{\mathrm{asymp}}(r)=\sqrt{\bar{b}r}$;
        (c) The Newtonian (Keplerian) curve, $v_{\mathrm{Kepler}}(r)= {\sqrt{\mu/r}}$;
        (d) The velocity minimum, $v^{(4)}_{m}=\left( 2\mu^2 \bar{b}^2 + \frac{10}{3}\mu^2\bar{a}^2 \right)^{1/8}$, located at radius $r^{(4)}_{m}=\sqrt{{\mu}/{\bar{b}}}$;
        (e) The line representing the minimum velocity plus 5\%, $v^{(4)}_m + 5\%$;
        (f) The flat rotation curve band, with a 10\% error margin;
        (g) The orbital velocity curve from the 2nd moment, $v(r)$.}
        }
\end{figure}

\newcommand{\oldfigc}{
\begin{figure}[t]
\centering
        \includegraphics[width=0.8\linewidth]{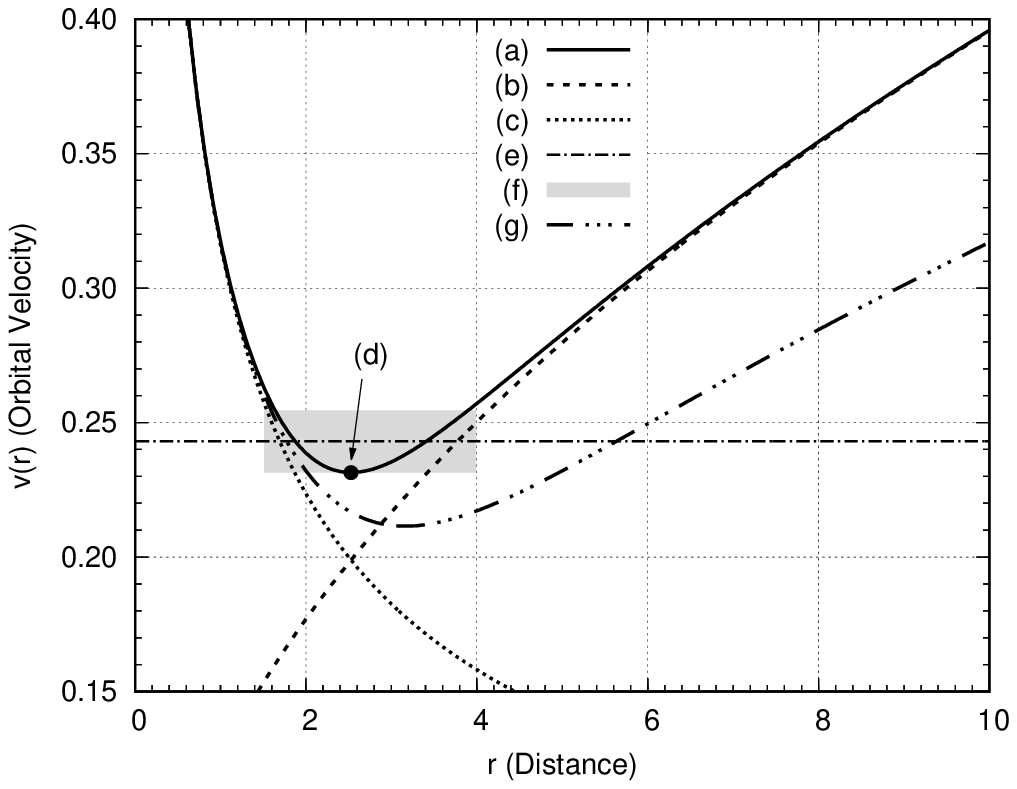}
        \label{fig:vorb4}
        \caption{{Rotation curve $v^{(4)}(r)$ in $\Phi^{(4)}$ 
        for ${\mu=0.1}$, ${\abar=0.01}$:
        (a)~$v^{(4)}(r)$;
        (b)~asymptote $v_{\mathrm{asymp}}(r)=\sqrt{\bbar r}$;
        (c)~$v_{\mathrm{Kepler}}(r)= {\sqrt{\mu/r}}$;
        (d)~minimum $v^{(4)}_{m}=\left( 2\mu^2 \bar{b}^2 + \frac{10}{3}\mu^2\bar{a}^2 \right)^{1/8}$ at  $r^{(4)}_{m}=\sqrt{{\mu}/{\bbar}}$;
        (e)~$v^{(4)}_m$ + 5\%;
        (f)~flat rotation curve with 10\% error margin;
        (g)~$v(r)$ in $\Phi^{(2)}$.}
        }
\end{figure}
}

\subsection{MOND from the 4th moment of the spin connection distribution}

Let us explore whether a version of MOND can be derived from the fourth-order qMOND dynamics in (\ref{ddot4}).  
Following the procedure outlined in Section \ref{sec311}, let us introduce a fictitious force due to the global mean-field acceleration $\bbar$, which results from the 
 fourth moment calculated in (\ref{avom4}):
\beq \label{ab4} 
\frac{GM}{r^2} = |\ddot{\bx}| - \bbar . 
\eeq
This definition is motivated by the large-distance linear asymptote of the potential, $\Phi^{(4)}(r) \rightarrow \bbar r$.

By defining the effective acceleration as $g= |\ddot{\mathbf{x}}| -\bbar$ and by interchanging the roles of force and acceleration in equation (\ref{ddot4}), we obtain a modified dynamics in the form:
\beq \label{mondo4}
\frac{GM}{r^2} = \left ( g^4 + \frac{10}{3} \abar^2 g^2  + \bbar{}^4 \right)^{1/4} - \bbar \, .
\eeq
This expression can be cast in the canonical MOND form,  Eq. (\ref{mond0}), with the interpolating function 
(IF), a function of the variable $x=g/a_0$,  given by:
\beq \label{if4}
\mu(x) = \mu^{(4)} (x)
 = \frac{5}{6^{3/2}x} \left(    \sqrt[4]{\frac{6^6}{5^4} x^4 + \frac{24\sqrt{6}}{5} x^2 + 1 }              
 - 1 \right) 
\eeq 
or, approximately, 
\beq \label{if4num}  
\mu^{(4)} (x)
 \approx \left(1 + \frac{{0.1575}}{x^2} + \frac{{0.0134}}{x^4}\right)^{1/4} - \frac{(0.0134)^{1/4}}{x}  \, . 
\eeq 
 The MOND acceleration constant $a_0$ 
is now defined in terms of $\bbar = \sqrt[4]{6}\abar$ and $\abar$ as follows: 
\beq \label{a04}
a_0 = \frac{6\bbar^3}{5\abar^2} = \frac{6^{7/4}}{5} \abar \approx 4.6\, \abar  \,.
\eeq 
Note that the coefficient in the last term of (\ref{if4num}) is fine-tuned to the value $(0.0134)^{1/4} \approx 0.3402328$ to prevent violation of the correct asymptotic behavior at $x \to 0^+$ due to the use of approximate numerical coefficients in the expression.


\begin{figure}[t]
\centering
        \includegraphics[width=0.8\linewidth]{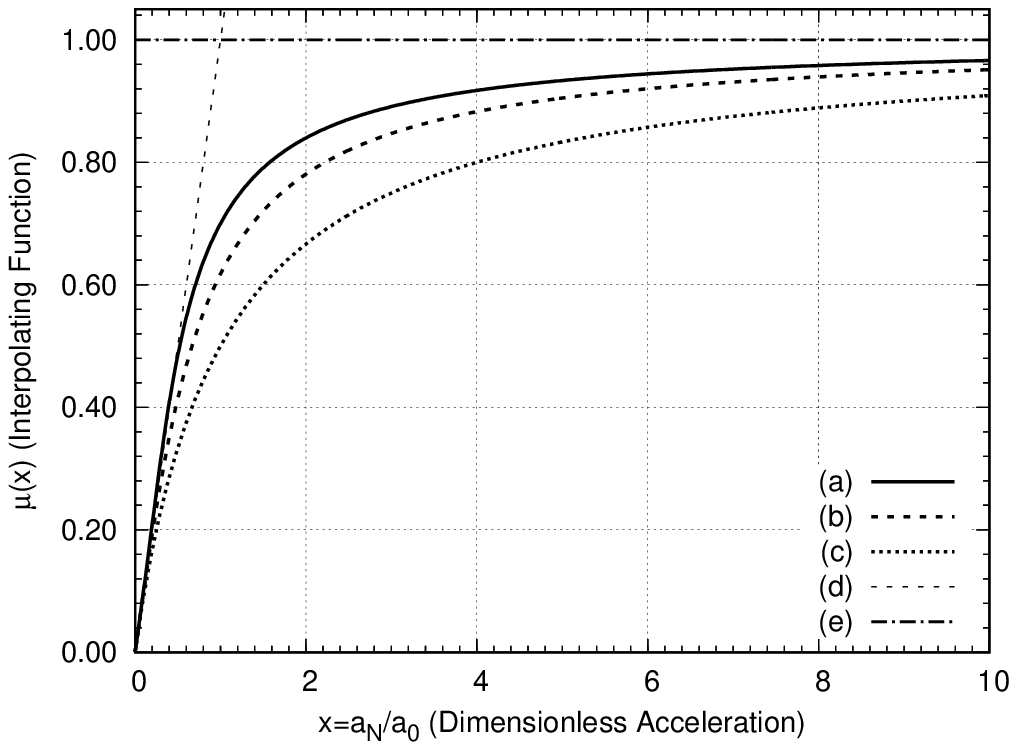}
        \label{fig:if4}
        \caption{{Comparison of MOND interpolating functions: 
(a) The interpolating function derived from the 4th moment, $\mu^{(4)}(x)$;
        (b) The interpolating function from the 2nd moment (previous result), $\mu^{(2)}(x)$;
        (c) The simple IF commonly used in MOND phenomenology, $\mu_{\mathrm{simple}}={x}/{(1+x)}$;
        (d) The deep-MOND asymptote ($\mu \sim x$ for $x \ll 1$);
        (e) The Newtonian asymptote ($\mu \sim 1$ for $x \gg 1$).   
        }        }
\end{figure}

A comparison with the interpolating function $\mu^{(2)}$ (derived from the variance of the geodesic equation in $(\ref{mux})$) shows that $\mu^{(4)}$ transitions to the Newtonian regime ($g \gg a_0$) slightly faster (cf. Fig. 4). This indicates that the sharper interpolating functions -- which are claimed to be necessary to mitigate potentially excessive MOND effects in the inner solar system\cite{solar1,solar2,solar3,solar4,solar5} -- could arise from the effects of higher moments of the fluctuating spin connection. The need to include the effects of higher moments could be related to the weak mass-multipole structure of planetary systems.


\subsection[qMOND and a generalized MOND from the 3rd moment \protect\newline of the spin connection distribution]{qMOND and a generalized MOND from the 3rd moment  \protect\\of the spin connection distribution} 

Let us consider the consequences of the third moment 
of the geodesic equation with random connection (\ref{mo3a}). Let us recall that in probability theory the third moment of a random variable 
is a measure of the skewness of its probability distribution. By averaging the third degree of 
(\ref{mo3a}) we obtain 
\begin{align}  \label{mom3}
(\ddot{\bx})^2 \ddot{x}{}^i &= \left\langle \left(\frac{\mu^2}{r^4} +2\mu \frac{x^j \omt^j}{r^3} +\omt^2\right) 
\left( - \mu \frac{x^i}{r^3} - \omt^i\right) \right\rangle \\
\label{mom3b}
& =  - \mu \frac{x^i}{r^3} \left (\frac{\mu^2}{r^4} + \frac{5}{3}\abar^2 \right) \ , 
\end{align} 
where we use the average values  
\beq \label{momij}
\langle \omt^i \omt^j \rangle = \frac{1}{3} \abar^2 \delta^{ij}  
\eeq
which are straightforward to calculate. 
Then, from (\ref{mom3}), we can calculate the 
absolute value of the acceleration: 
\beq \label{ddx3}
|\ddot{\bx}| =  \left (\frac{\mu^3}{r^6} +  \frac{5}{3}\abar^2 \frac{\mu}{r^2} \right)^{1/3} \, , 
\eeq
which can be written in the vector form (\ref{vect}) where the potential derived from (\ref{ddx3}) is given by 
\beq \label{phi3}
\Phi^{(3)}(r) = - \frac{\mu}{r}\, {}_2F_1\left(-\frac{1}{3}, -\frac{1}{4}; \frac{3}{4}; -\frac{5 \abar^2 r^4}{3 \mu^2}\right) \, . 
\eeq
This potential {(cf. Fig. 5)} approaches the Newtonian limit ${-\mu/r}$ for small $r$, and for large $r$, it grows asymptotically as $\sqrt[3]{45 \mu\abar^2}\, r^{1/3}$. 

Figure 6 compares the qMOND potentials $\Phi^{(2)}(r)$, $\Phi^{(3)}(r)$, $\Phi^{(4)}(r)$, and the Newtonian $-\mu/r$. 

\begin{figure}[t]
\centering
        \includegraphics[width=0.8\linewidth]{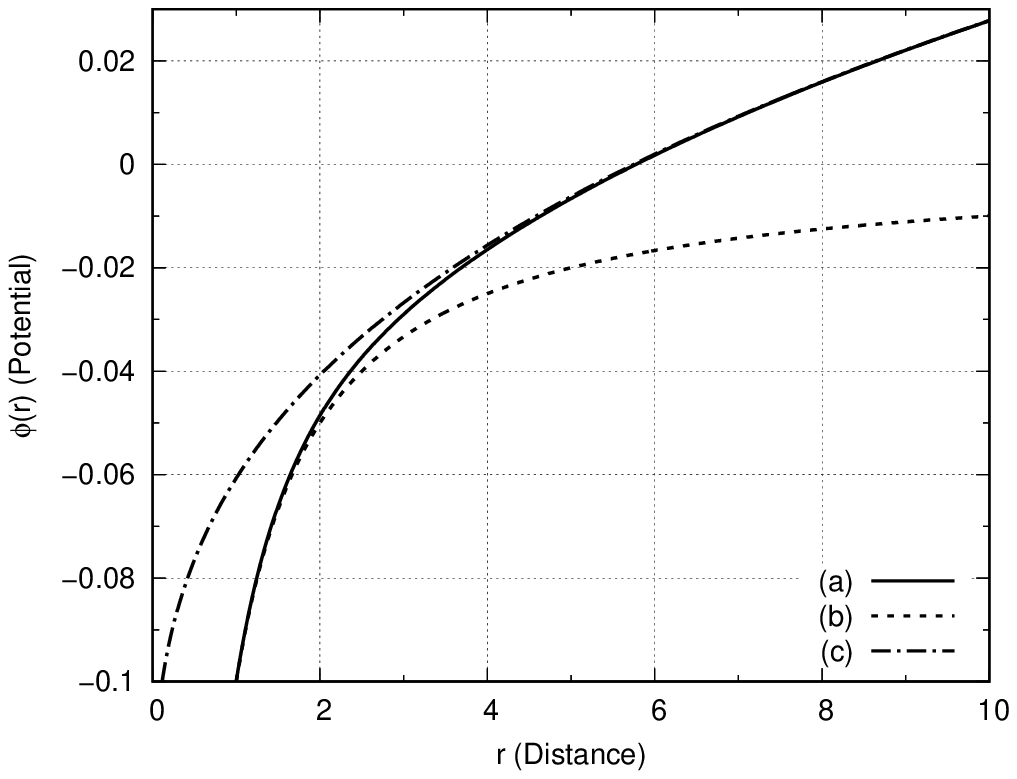}
        \label{fig:phi3}
        \caption{
        qMOND potential $\Phi^{(3)}(r)$ for ${\mu=0.1}$, ${\abar=0.01}$: 
        (a) 
        $\Phi^{(3)}(r)$;
        (b) 
        $\Phi_{\mathrm{Newton}}(r)=-\frac{\mu}{r}$; 
        (c)~the  asymptote at $r\rightarrow \infty$: $\Phi^{(3)}_{\rm{asymp}}(r) \rightarrow \sqrt[3]{45 \mu \bar{a}^2 r} - \frac{\sqrt[4]{\frac{5}{3}} \Gamma \left(-\frac{1}{12}\right) \Gamma \left(\frac{3}{4}\right) \sqrt{\mu  \bar{a}}}{\Gamma \left(-\frac{1}{3}\right)}$.
        }   
\end{figure}

\begin{figure}[t]
\centering
        \includegraphics[width=0.8\linewidth]{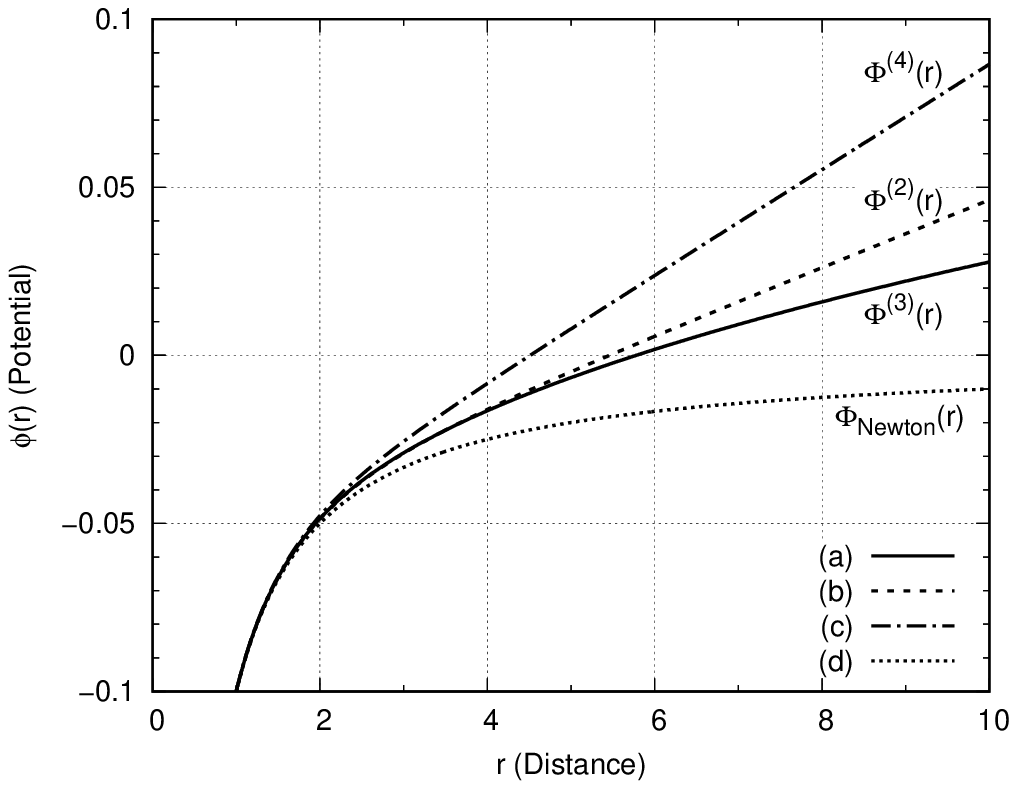}
        \label{fig:phi234}
        \caption{
        {Comparison of qMOND potentials $\Phi^{(3)}(r)$, $\Phi^{(2)}(r)$, $\Phi^{(4)}(r)$ with the Newtonian potential, ${\mu=0.1}$, ${\abar=0.01}$: 
        (a) 
        $\Phi^{(3)}(r)$;
        (b) 
        $\Phi^{(2)}(r)$;
        (c) 
         $\Phi^{(4)}(r)$;
        (d) 
        $\Phi_{\mathrm{Newton}}(r)=-{\mu}/{r}$.}
        }
\end{figure}

\subsection{The rotation curve in $\boldsymbol{\Phi^{(3)}}$ potential} 

For the orbital velocity we now obtain: 
\beq  \label{v3}  
|\mathbf{v}|{}^{(3)} = \left (\frac{\mu^3}{r^3} +  \frac{5}{3}\abar^2 {\mu}{r} \right)^{1/6} \, . 
\eeq
This orbital velocity has the minimal value 
\beq \label{vm3}
v_m^{(3)} = 2^{1/3} 5^{1/8} 3^{-1/4} \left(\mu \bar{a})^{1/4} \approx 1.17 (\mu \bar{a} \right)^{1/4} 
 \, , 
\eeq
which occurs at the distance $r_m$ from the central mass: 
\beq \label{rm3}
r_m^{(3)} =  \left(\frac{9\mu^2}{5\bar{a}^2}\right)^{1/4}  \approx 1.16 \left(\frac{\mu^2}{\bar{a}^2}\right)^{1/4} 
\, . 
\eeq
Around $r_m$, the orbital velocity again behaves as a very  flat parabola 
\begin{align}\label{v3flat}
\begin{split}
|\mathbf{v}|{}^{(3)} &\approx 
 1.17\, (\mu \bar{a})^{1/4} + \frac{v_m}{4r_m^2} (r-r_m)^2   
\\ 
& \approx 1.17\, (\mu \bar{a})^{1/4} + 
 0.22\, (\abar^5\!/\mu^3)^{1/4} (r-r_m)^2 \, . 
 \end{split} 
\end{align}
The comparison with the Taylor expansions in (\ref{vr2}) and (\ref{v4flat}) reveals that 
the rotation curve derived from the third moment is $0.59/0.22 \approx 2.7$ times 
flatter than the curve derived from the second moment and $1.21/0.22  \approx 5.5$ times 
flatter than the curve derived from the fourth moment. 

The $\propto r^{1/6}$ slow asymptotic growth of the orbital velocity in (\ref{v3}) widens the region of the approximately flat rotation curve relative to the qMOND 
potentials $\Phi^{(2)}(r)$ and $\Phi^{(4)}(r)$. It is thus expected that the qMOND derived from higher odd moments of (\ref{mo3a}) will yield even flatter asymptotic rotation curves, approaching the limiting flat rotation curve, $\propto r^{0}$ (or $\propto r^{1/\infty}$), in the limit of infinite odd moments of 
geodesic equation (\ref{mo3a}).

%
%

Figure 7 displays the rotation curve  $v^{(3)}(r)$ and compares it with $v(r)$, $v^{(4)}(r)$, and the Keplerian rotation curve. 

\begin{figure}[t]
\centering
        \includegraphics[width=0.8\linewidth]{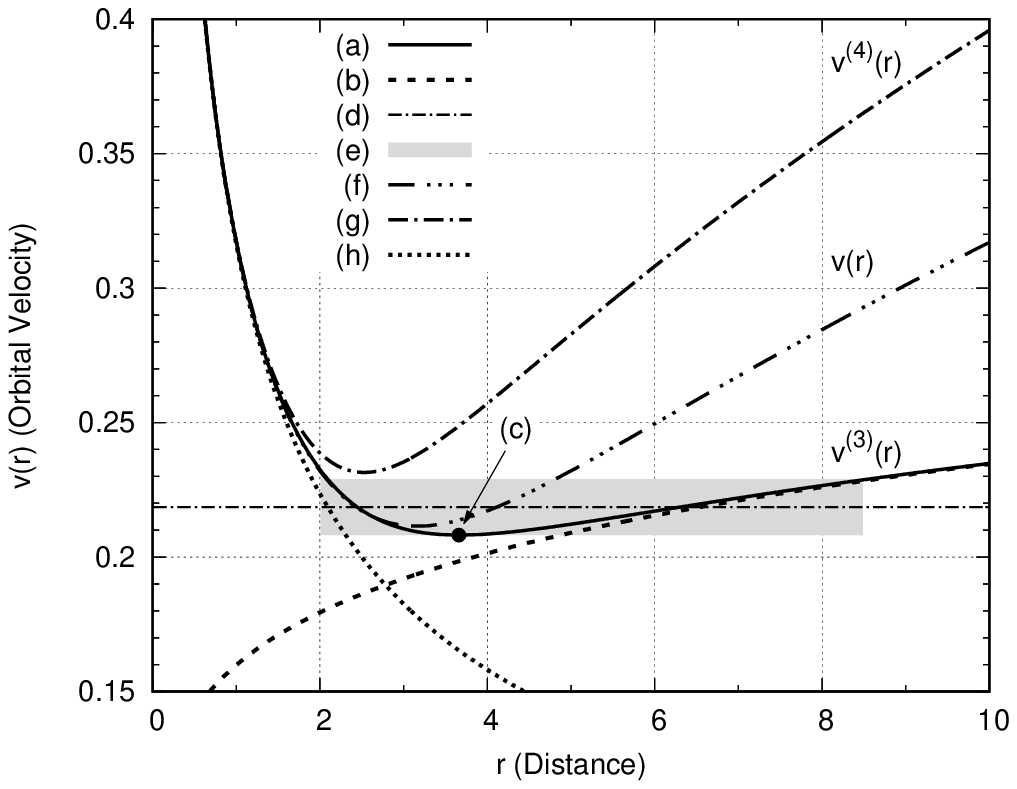}
        \label{fig:vorb3}
        \caption{{Comparison of qMOND rotation curves} $v^{(3)}(r)$, $v^{(4)}(r)$, and $v(r)$ for parameters $\mu=0.1$ and $\bar{a}=0.01$:
            (a) The orbital velocity curve for the 3rd moment, $v^{(3)}(r)$;
            (b) The asymptote of the $v^{(3)}$ curve: $v^{(3)}_{\rm{asympt}}(r)=\sqrt[6]{{5}\mu r \bar{a}^2/3}$;
            (c) The minimum of the $v^{(3)}$ orbital velocity, $v^{(3)}_{m}=\sqrt[3]{2} \sqrt[8]{5 \mu^2 \bar{a}^2/9}$, located at $r^{(3)}_{m}=\sqrt[4]{{9 \mu^2}/{5 \bar{a}^2}}$;
           (d)  The line representing $v^{(3)}_m$ plus 5\%;
           (e) The flat rotation curve band derived from $v^{(3)}$, with a 10\% error margin;
            (f)  The orbital velocity curve for the 2nd moment, $v(r)$;
            (g) The orbital velocity curve for the 4th moment, $v^{(4)}(r)$;
            (h) The Newtonian (Keplerian) orbital velocity curve: $v_{\mathrm{Kepler}}(r)=\sqrt{{\mu}/{r}}$.
            }
\end{figure}
\newcommand{\oldfigb}{
\begin{figure}[t]
\centering
        \includegraphics[width=0.8\linewidth]{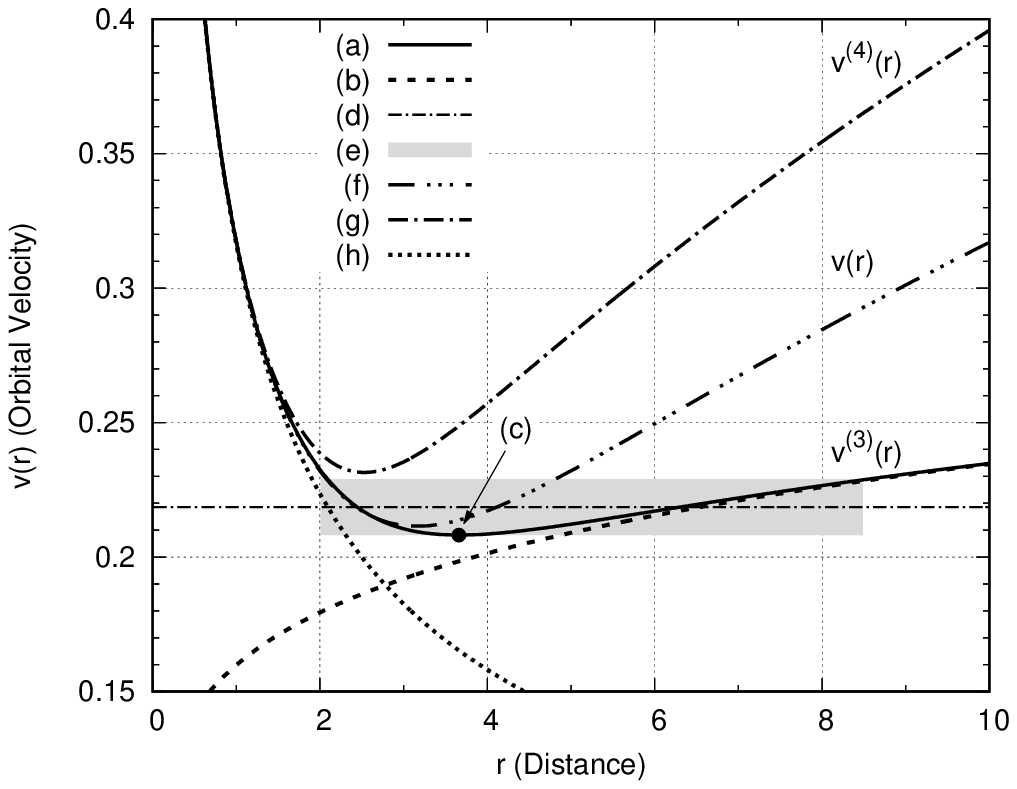}
        \label{fig:vorb3}
        \caption{\JK{Orbital velocity $v^{(3)}(r)$ compared with $v^{(4)}(r)$ and $v(r)$, for ${\mu=0.1}$, ${\abar=0.01}$: 
        (a) orbital velocity curve for the 3rd moment: $v^{(3)}(r)$; 
        (b) the asymptote of the qMOND orbital velocity: $v^{(3)}_{\rm{asympt}}(r)=\sqrt[6]{{5}\mu  r \bar{a}^2/3}$; 
        (c) minimum of the qMOND orbital velocity: $v^{(3)}_{m}=\sqrt[3]{2} \sqrt[8]{5 \mu^2 \bar{a}^2/9}$ at ${r^{(3)}_{m}=\sqrt[4]{{9 \mu^2}/{5 \bar{a}^2}}}$;
        (d) $v^{(3)}_m$ + 5\% line;  
        (e) flat rotation curve from $v^{(3)}$ with 10\% error margin; 
        (f) orbital velocity curve for the 2nd moment: $v(r)$; 
        (g) orbital velocity curve for the 4th moment: $v^{(4)}(r)$; 
        (h) the Keplerian orbital velocity curve: $v_{Kepler}(r)=\sqrt{{\mu}/{r}}$.}
        }
\end{figure}
}

\subsubsection{The emergence of a modified MOND (mMOND)}

As the asymptotic behavior of $\Phi^{(3)}$ at large $r$ is $\propto r^{1/3}$ rather than linear (as in the case of $\Phi^{(2)}$ and $\Phi^{(4)}$), the analogue of the mean-field acceleration in this case is $r$-dependent. 
It is also evident from (\ref{ddx3}) where the asymptotic acceleration due to the fluctuations 
(quantum randomness) of spin connection in SCF 
equals 
\beq \label{abbar}
\abbar(r) =  \sqrt[3]{\frac{5\abar^2\mu}{3}}\, r^{-2/3}. 
\eeq 
Following our previous scheme of obtaining MOND and its theoretical interpolation function, let us  
transform the dynamics to the non-inertial reference frame of asymptotic accelerations $\abbar(r)$ 
by adding a fictitious force to the Newtonian force $\mu/r^2$ (c.f. (\ref{aa2}) and (\ref{ab4})): 
\beq \label{ddotx3}
|\ddot{x}| - \abbar(r) = \frac{\mu}{r^2} \, .
\eeq
By substituting this redefinition of force and acceleration into the third order qMOND law in (\ref{ddx3}) 
and introducing new Newtonian acceleration in accordance with (\ref{ddotx3}): 
\beq \label{g3}
g(r) = |\ddot{x}| - \abbar(r) \, , 
\eeq 
we obtain 
\beq \label{mmond3}
\frac{\mu}{r^2} = \sqrt[3]{ g^3(r) + \abbar(r){}^3} - \abbar(r)  \, . 
\eeq
Let us try to cast this modified dynamics in a MOND-like form 
\beq
\frac{\mu}{r^2} = \bar{\bar{\mu}} \left(\frac{g(r)}{a_0(r)}\right) g(r) \, , 
\eeq
with an $r-$dependent transition function $\mubbar (r)$
 and a properly defined $r-$dependent acceleration scale $a_0(r)$. The transition function $\bar{\bar{\mu}}(r)$ is supposed to interpolate between 
the Newtonian dynamics at large accelerations $g(r) \gg a_0(r)$ and a ``deep-MOND"-like dynamics 
at small accelerations $g(r)$. 


As the asymptote of the function $\bar{\bar{\mu}}$  at 
$g \gg \abbar$ is $ \bar{\bar{\mu}} \approx (1 - {\abbar}/{g})$ and the asymptote at 
$g \ll \abbar$ is $ \bar{\bar{\mu}} \approx \frac{1}{3} {g^2}/{\abbar^2}$, 
 we conclude that by identifying  
\beq \label{a0r}
a_0(r) = \sqrt{3}\, \abbar(r) \, , 
\eeq
 we can write $\bar{\bar{\mu}}$ as a function of $u(r) = g(r)/a_0(r)$, 
 which is itself a function of $r$ ($ u(r) \propto r^{-4/3}$), namely: 
\begin{align} \label{ifmmond} 
\begin{split}
\bar{\bar{\mu}} (u) &= \sqrt[3]{ 1+ ({a_0}^{3}/3^{3/2})g^{-3}} - a_0 (\sqrt{3}g)^{-1} 
 \\ 
& = \left(1 + \frac{1}{3^{3/2}u^3}\right)^{1/3} - \frac{1}{\sqrt{3}u} \, , 
\end{split}
\end{align}
which behaves as $\sim\! \left(1 - {1}/{\sqrt{3}u}\right)$ at large $u$ and $\sim\! u^2$ at small $u$. This interpolates between the Newtonian dynamics at large $g/a_0$ and a new ``deep modified MOND" (or ``deep-mMOND") regime given by: 
\beq \label{dmmond}
\frac{g(r)^3}{a_0(r)^2} = \frac{\mu}{r^2} . 
\eeq

It is interesting to find the asymptotic rotation curve originating from the mMOND. 
Let us recall that Milgrom's deep-MOND law 
\beq
\frac{g^2}{a_0} = \frac{\mu}{r^2}
\eeq
automatically leads to the flat rotation curve and the baryonic Tully-Fischer relation $v^4 = \mu a_0$,\cite{tf1,tf2,tf3} which was a theoretical prediction of MOND back in 1983 \cite{mond1,mond2,mond3,mond4}. 
By substituting the kinematic relation $g=v^2/r $ and the $r$-dependent $a_0(r)$ from  
(\ref{a0r}) and (\ref{abbar})  into (\ref{dmmond}), we obtain 
\beq
v^6 = \sqrt[3]{\frac{75 \abar^4 \mu^5}{r}}  \, , 
\eeq
that is, the asymptotic orbital rotation velocity decays very slowly as $\propto r^{-1/18}$ 
at distances $r>r_\mathrm{mmond}$  where the effects of $r$-dependent Milgromian acceleration scale 
$a_0(r)$ defined in (\ref{a0r}) dominate over the Newtonian acceleration $g= \mu/r^2$. 
It is easy to find that $a_0(r) > \mu/r^2$ at distances larger than 
\beq
r_\mathrm{mmond} = \frac{1}{\sqrt[4]{5\sqrt{3}}}\sqrt{\frac{\mu}{\abar}} \approx 0.58\, \sqrt{\frac{\mu}{\abar}} 
\, .
\eeq 
As this critical distance is close to the Milgromian distance in MOND, 
\beq 
r_\mathrm{mond} = \sqrt{\frac{\mu}{a_0}} \, , 
\eeq 
because the numerical value of Milgrom's constant $a_0$ is of the same order of magnitude as the 
value of $\abar$ (cf., for example, equation (\ref{g0})), 
we can expect the effects of mMOND to be manifested at similar  galactic scales as MOND, 
rather than on the larger scales of galactic clusters or superclusters, as one could have assumed {\em a priori}. Namely, if Milgrom's $a_0$ is identified with $2\abar$, as in (35), then 
$r_\mathrm{mmond} \approx 
0.82\, r_\mathrm{mond} $, and if $a_0$ is identified with $\sqrt{6} \ \bar{b}$, as in (\ref{a04}), 
then $r_\mathrm{mmond} \approx 
1.14\,  r_\mathrm{mond} $. 

Figures 8 and 9 compare the qMOND interpolating function ${\mu^{(3)}(x)}$ against the IFs derived from the second and fourth moments (${\mu^{(2)}(x)}$ and ${\mu^{(4)}(x)}$), as well as the simple interpolating function of MOND.
 

\begin{figure}[t]
\centering
        \includegraphics[width=0.8\linewidth]{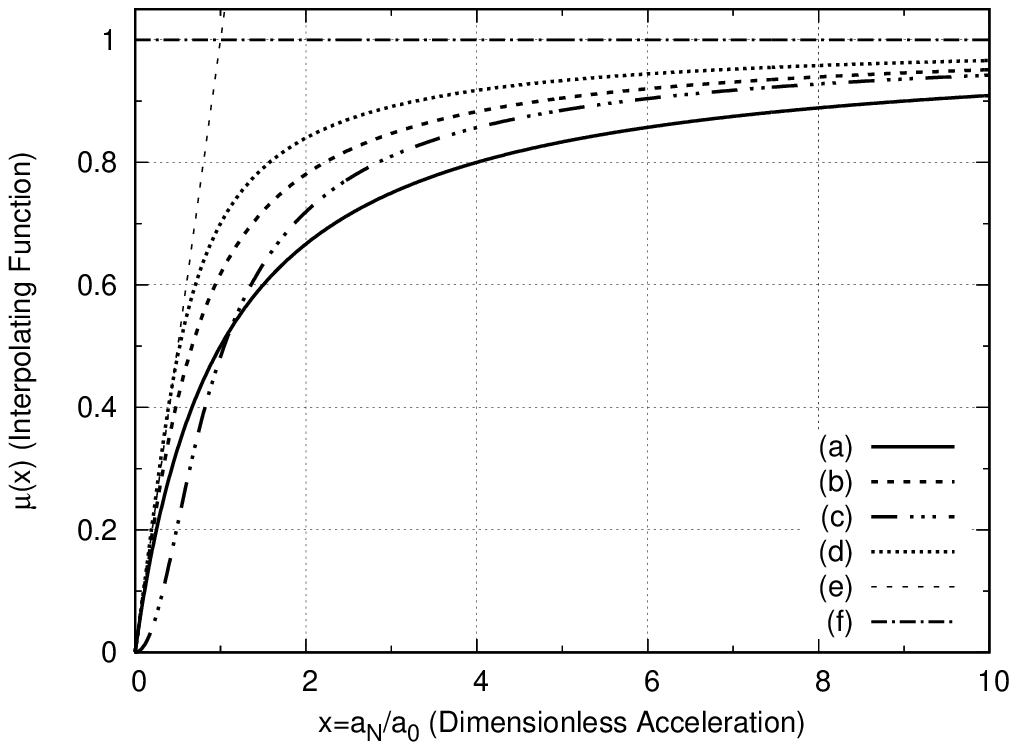}
        \label{fig:if234}
        \caption{{Comparison of MOND and mMOND interpolating functions: \\
        (a) simple $\mu=\frac{x}{1+x}$;
        (b) $\mu^{(2)}(x)$; 
        (c) $\mu^{(3)}(x)$ (mMOND);
        (d) $\mu^{(4)}(x)$;
        (e) the deep-MOND asymptote ($\mu \propto x$ at $x \ll 1$);
        (f) the Newtonian asymptote ($\mu \rightarrow 1$ at $x \gg 1$).}
        }
\end{figure}

\begin{figure}[h!]
\centering
        \includegraphics[width=0.8\linewidth]{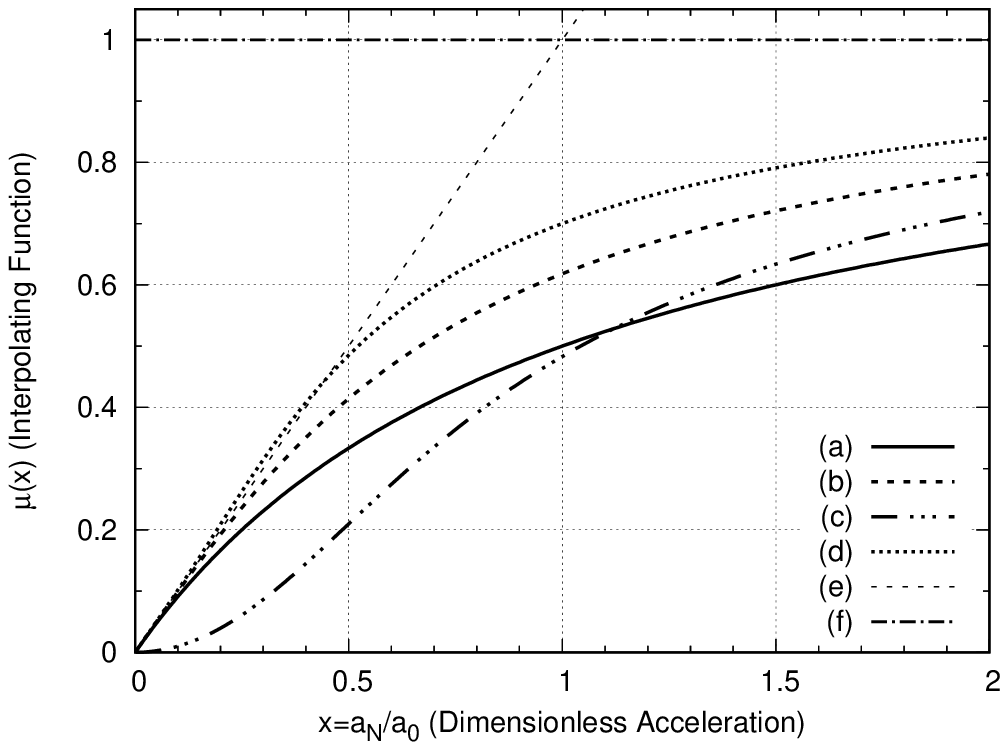}
        \label{fig:if234zoom}
        \caption{
        {Comparison of MOND and mMOND IFs in the deep non-Newtonian regime:\\
        (a) simple $\mu=\frac{x}{1+x}$;
        (b) $\mu^{(2)}(x)$;
        (c) $\mu^{(3)}(x)$ (mMOND);
        (d) $\mu^{(4)}(x)$; 
        (e) the deep-MOND asymptote ($\mu \propto x$ at $x \ll 1$);
        (f) the Newtonian asymptote ($\mu \rightarrow 1$ at $x \gg 1$).}
        }
\end{figure}


\section{Conclusions}  

Prior work \cite{epl25,mg17,dice24,bial22} demonstrated that Milgrom's modification of Newtonian dynamics (MOND) \cite{mond1,mond2,mond3,mond4} -- posited as an alternative to dark matter -- 
can be derived within the framework of precanonical quantum gravity (pQG) \cite{ikv1,ikv2,ikv3,ikv4,ikv5,ik-mink} from the second moment of the non-relativistic geodesic equation with a random spin connection term.
This equation models a test particle interacting with both the gravitating mass and the random connection which arises from quantum fluctuations in the spin connection foam, the vacuum state of quantum gravity according to pQG. The derivation proceeds by first obtaining a modified Newtonian potential (the qMOND potential) from the SCF fluctuations and then performing a transformation to the non-inertial reference frame of the mean-field acceleration ($\bar{a}$) originating from these fluctuations. This mean-field acceleration, derived within pQG, was found to be roughly compatible with the phenomenological value of the Milgrom acceleration $a_0$, as first noticed in [14].  


This paper extends the previous work by exploring the consequences of the higher-order moments (third and fourth) of the same geodesic equation ($\ref{mo3a}$), which includes the effects of random spin connection arising from the quantum gravitational fluctuations of the SCF. These fluctuations lead to specific asymptotic behaviors of the modified gravitational potentials derived from the different moments: The second and fourth moments yield potentials ($\Phi^{(2)}$ and $\Phi^{(4)}$, respectively) with linearly growing asymptotes ($\propto r$) at large distances. The asymptotic behaviors of these potentials, ${\Phi^{(2)} \propto \abar r}$ and ${\Phi^{(4)} \propto \bbar r}$, are controlled by the standard deviation ($\abar$) and the fourth root of the fourth moment ($\bbar$), respectively. These parameters originate from the spin connection distribution derived from the precanonical wave function of the spin connection foam. Crucially, the asymptotic slope derived from the fourth moment is higher than that derived from the variance (or the standard deviation), a relationship explicitly defined by $\bbar = \sqrt[4]{6}\abar$.
 %
 Conversely, the third moment yields a potential with a power-law asymptotic growth: $\Phi^{(3)} \propto r^{1/3}$. We present the explicit expressions of these potentials in terms of the Gauss hypergeometric functions ${}_2F_1$ (for $\Phi^{(2)}$ and $\Phi^{(3)}$) and the Appell functions $F_1$ for $\Phi^{(4)}$. 
More general 
universal expressions in terms of Lauricella functions, a multivariate generalization of the hypergeometric series, are currently a work in progress.\cite{inprep-l}

The resulting orbital velocities in the modified potentials $\Phi^{(2)}$ and $\Phi^{(4)}$ are asymptotically growing ($\propto r^{1/2}$), which manifests as approximately flat rotation curves around their respective minima $r_m$ at $\sqrt{\mu/\abar}$ and $\sqrt{\mu/\overline{b}}$. The orbital velocity in the potential $\Phi^{(3)}$ grows asymptotically much slower ($\propto r^{1/6}$), so that it is approximately flat in a wider region around the farther situated minimum at $r_m = \sqrt{3\mu / \sqrt{5}\abar}$.  

When transformed to the non-inertial frames of the asymptotic mean-field ac\-cel\-erations derived from the asymptotes of the qMOND potentials, we recover the Milgromian MOND behavior from the $\Phi^{(2)}$ and $\Phi^{(4)}$ potentials. This yields theoretically derived interpolating functions  
and the relation between the Milgrom acceleration $a_0$ and the characteristics of the spin connection distribution function derived from pQG: standard deviation $\abar$ and 
the fourth root of the fourth moment~$\bbar$.  

In the case of the dynamics governed by the $\Phi^{(3)}$ potential, we similarly derive a modification of MOND ($\text{mMOND}$) and the corresponding $r$-dependent interpolating function. This function interpolates between the Newtonian and the deep-$\text{mMOND}$ regime $g^3/a_0^2 = GM/r^2$ with an $r$-dependent $a_0(r) \propto r^{-2/3}$. This modification leads to an almost flat rotation curve ($\propto r^{-1/18}$) at approximately the same scales where canonical $\text{MOND}$ leads to the exactly flat rotation curves.

Let us note that the theoretical IFs derived in $(\ref{mux})$, $(\ref{if4})$ and $(\ref{ifmmond})$
are defined uniquely only under specific underlying physical assumptions. 
 These assumptions are applied on top of the point central mass approximation, the non-relativistic static approximation of the Minkowskian SCF, and the related instantaneous gravitational action at a distance, and include:
\begin{itemize} 
\item The lack of back-reaction of the central mass $M$ on the state of the SCF. 
\item The central mass remaining rigidly fixed at the origin, despite the quantum fluctuations of spin connection influencing both the test particle and the central body equally -- as required by the Equivalence Principle.
\item The neglect of long-distance correlations in the non-relativistic SCF. These correlations follow from the two-point solutions of (\ref{mo20p}) and may lead to an additional correlational contribution to the force between the test particle and the central mass -- again, due to the Equivalence Principle.
\end{itemize}
The work on lifting those assumptions is currently in progress. Instead, this paper concentrates on the theoretical derivation of new interpolating functions from the higher-order moments of equation ($\ref{mo3a}$). %
It remains to be understood how the physical characteristics of a gravitational system determine the necessary order of the moment (2nd, 3rd, 4th, or higher) required for the correct description of the system applicable at large distances and small accelerations, where the effects of quantum randomness of the spin connection manifest themselves.

Moreover, it is both mysterious and worrisome that the higher even moments lead to increasing slopes of the linear asymptotes of the corresponding qMOND potentials at large distances (cf. $(\ref{avom2k})$). This appears to be an unphysical behaviour in the corresponding qMOND regimes, though it is corrected by transitioning to the MOND regime. On the other hand, the higher odd moments lead to increasingly flat rotation curves in both the qMOND and mMOND regimes.

Despite those uncertainties, the paper demonstrates that the underlying quantum fluctuations of the spin connection, as described by pQG and analyzed via higher-order moments, can account for various phenomena usually attributed to Milgrom's phenomenological MOND. This result provides a first-principles quantum-gravitational origin for the non-Keplerian behavior observed in large-scale galactic dynamics, achieving this without superficially trying out different modifications of classical General Relativity or invoking yet unobserved dark matter candidates.

\section*{Acknowledgments} 
Authors thank Natascha Riahi for her thoughtful comments on a pre-final version of the manuscript. 
IK and JK thank Marek Czachor for his insightful question regarding the MOND--precanonical quantum gravity 
correspondence, which has prompted us to consider higher moments in this paper.
IK acknowledges stimulating discussions with Indranil Banik about the viability of the results of previous papers [1--3] in the Solar system and Saturn system, and the Organizers of\, ``My Favorite Dark Matter Model'' 
on the Azores (Portugal) in April 2025 for the opportunity to present the results of an earlier stage of this work.







\end{document}